\documentclass[]{aastex63}

\shorttitle{switchback}
\shortauthors{Huang et al.}
\graphicspath{{./}{figures/}}

\begin{document}

\title{The Structure and Origin of Switchbacks: Parker Solar Probe Observations}

\correspondingauthor{Jia Huang}
\email{huangjia.sky@gmail.com}

\author[0000-0002-9954-4707]{Jia Huang}
\affiliation{Space Sciences Laboratory, University of California, Berkeley, CA 94720, USA.}

\author[0000-0002-7077-930X]{J. C. Kasper}
\affiliation{BWX Technologies, Inc., Washington DC 20001, USA.}
\affiliation{Climate and Space Sciences and Engineering, University of Michigan, Ann Arbor, MI 48109, USA.}

\author[0000-0002-0646-2279]{L. A. Fisk}
\affiliation{Climate and Space Sciences and Engineering, University of Michigan, Ann Arbor, MI 48109, USA.}

\author[0000-0001-5030-6030]{Davin E. Larson}
\affiliation{Space Sciences Laboratory, University of California, Berkeley, CA 94720, USA.}

\author[0000-0001-6077-4145]{Michael D. McManus}
\affiliation{Space Sciences Laboratory, University of California, Berkeley, CA 94720, USA.}

\author[0000-0003-4529-3620]{C. H. K. Chen}
\affiliation{School of Physics and Astronomy, Queen Mary University of London, London E1 4NS, UK.}

\author[0000-0002-7365-0472]{Mihailo M. Martinovi\'c}
\affiliation{Lunar and Planetary Laboratory, University of Arizona, Tucson, AZ 85719, USA.}
\affil{LESIA, Observatoire de Paris, Université PSL, CNRS, Sorbonne Université, Université de Paris, 5 place Jules Janssen, 92195 Meudon, France.}

\author[0000-0001-6038-1923]{K. G. Klein}
\affiliation{Lunar and Planetary Laboratory, University of Arizona, Tucson, AZ 85719, USA.}

\author{Luke Thomas}
\affiliation{Climate and Space Sciences and Engineering, University of Michigan, Ann Arbor, MI 48109, USA.}

\author[0000-0003-2981-0544]{Mingzhe Liu}
\affil{LESIA, Observatoire de Paris, Université PSL, CNRS, Sorbonne Université, Université de Paris, 5 place Jules Janssen, 92195 Meudon, France.}

\author[0000-0002-2229-5618]{Bennett A. Maruca}
\affiliation{Department of Physics and Astronomy, University of Delaware, Newark, DE 19716, USA.}
\affil{Bartol Research Institute, University of Delaware, Newark, DE 19716.}

\author[0000-0002-4299-0490]{Lingling Zhao}
\affil{Department of Space Science and CSPAR, The University of Alabama in Huntsville, Huntsville, AL 35805, USA.}

\author[0000-0002-0065-7622]{Yu Chen}
\affil{Department of Space Science and CSPAR, The University of Alabama in Huntsville, Huntsville, AL 35805, USA.}

\author[0000-0002-7570-2301]{Qiang Hu}
\affil{Department of Space Science and CSPAR, The University of Alabama in Huntsville, Huntsville, AL 35805, USA.}

\author[0000-0002-6849-5527]{Lan K. Jian}
\affiliation{Heliophysics Science Division, NASA Goddard Space Flight Center, Greenbelt, MD 20771, USA}

\author[0000-0003-1138-652X]{J. L. Verniero}
\affiliation{Heliophysics Science Division, NASA Goddard Space Flight Center, Greenbelt, MD 20771, USA}

\author[0000-0002-2381-3106]{Marco Velli}
\affiliation{Department of Earth, Planetary and Space Sciences, University of California, Los Angeles CA 90095, USA}

\author[0000-0002-0396-0547]{Roberto Livi}
\affiliation{Space Sciences Laboratory, University of California, Berkeley, CA 94720, USA.}

\author[0000-0002-7287-5098]{P. Whittlesey}
\affiliation{Space Sciences Laboratory, University of California, Berkeley, CA 94720, USA.}

\author[0000-0003-0519-6498]{Ali Rahmati} 
\affiliation{Space Sciences Laboratory, University of California, Berkeley, CA 94720, USA.}

\author[0000-0002-4559-2199]{Orlando Romeo}
\affiliation{Space Sciences Laboratory, University of California, Berkeley, CA 94720, USA.}

\author[0000-0001-6692-9187]{Tatiana Niembro}
\affiliation{Smithsonian Astrophysical Observatory, Cambridge, MA 02138 USA.}

\author[0000-0002-5699-090X]{Kristoff Paulson}
\affiliation{Smithsonian Astrophysical Observatory, Cambridge, MA 02138 USA.}

\author[0000-0002-7728-0085]{M. Stevens}
\affiliation{Smithsonian Astrophysical Observatory, Cambridge, MA 02138 USA.}

\author[0000-0002-3520-4041]{A. W. Case}
\affiliation{Smithsonian Astrophysical Observatory, Cambridge, MA 02138 USA.}

\author[0000-0002-1573-7457]{Marc Pulupa}
\affiliation{Space Sciences Laboratory, University of California, Berkeley, CA 94720, USA.}

\author[0000-0002-1989-3596]{Stuart D. Bale}
\affil{Physics Department, University of California, Berkeley, CA 94720-7300, USA.}
\affil{Space Sciences Laboratory, University of California, Berkeley, CA 94720, USA.}
\affil{The Blackett Laboratory, Imperial College London, London, SW7 2AZ, UK.}
\affil{School of Physics and Astronomy, Queen Mary University of London, London E1 4NS, UK.}

\author[0000-0001-5258-6128]{J. S. Halekas}
\affil{Department of Physics and Astronomy, University of Iowa, Iowa City, IA 52242, USA.}



\begin{abstract}

Switchbacks are rapid magnetic field reversals that last from seconds to hours. Current Parker Solar Probe (PSP) observations pose many open questions in regard to the nature of switchbacks. For example, are they stable as they propagate through the inner heliosphere, and how are they formed? In this work, we aim to investigate the structure and origin of switchbacks. In order to study the stability of switchbacks, we suppose the small-scale current sheets therein are generated by magnetic braiding, and they should work to stabilize the switchbacks. With more than one thousand switchbacks identified with PSP observations in seven encounters, we find many more current sheets inside than outside switchbacks, indicating that these microstructures should work to stabilize the S-shaped structures of switchbacks. Additionally, we study the helium variations to trace the switchbacks to their origins. We find both helium-rich and helium-poor populations in switchbacks, implying that the switchbacks could originate from both closed and open magnetic field regions in the Sun. Moreover, we observe that the alpha-proton differential speeds also show complex variations as compared to the local Alfvén speed. The joint distributions of both parameters show that low helium abundance together with low differential speed is the dominant state in switchbacks. The presence of small-scale current sheets in switchbacks along with the helium features are in line with the hypothesis that switchbacks could originate from the Sun via interchange reconnection process. However, other formation mechanisms are not excluded.

\end{abstract}



\section{Introduction} \label{sec:intro}
Parker Solar Probe (PSP) is the first mission to touch the Sun, and it aims to uncover the solar wind properties in the inner heliosphere \citep{Fox-2016}. The PSP has completed 14 orbits by 2022 December, and its deepest perihelion reached a radial distance of about 0.062 au or 13.3 solar radii ($\mathrm{R_S}$). 
One of the most extraordinary observations in the near-Sun environment is the prevalent presence of switchbacks, which are defined as the rapid magnetic field reversals that last from seconds to hours \citep[e.g.][]{Bale-2019, kasper-2019, de-2020, horbury-2020, Mozer-2020}. Many new properties of switchbacks are uncovered, which also bring new challenges to be explained.

The origin of switchbacks is a primary question to be answered and current studies indicate the switchbacks could form either in the interplanetary space or from the Sun. 

Several works support that switchbacks are formed in the interplanetary space. With the Helios observations from 0.3 au to 1 au, \citet{macneil-2020} find the total sampled time of switchbacks (durations larger than 40 s) increases with radial distances, implying the switchbacks could form in the interplanetary space through processes such as velocity shears, draping over ejecta, or waves and turbulence. Furthermore, \citet{squire-2020} and \citet{Mallet-2021} reveal that the switchbacks could form in-situ in the expanding solar wind based on the numerical study of the low-amplitude outward-propagating Alfvénic fluctuations. In addition, \citet{schwadron-2021} propose a method that the switchbacks could be evolved above the Alfvén point due to the so-called super-Parker spiral that is formed by the magnetic field footpoints walk from the source of slow wind to faster wind. Moreover, \citet{larosa-2021} demonstrate that there are Alfvénic, fast, and slow mode signatures in switchbacks, implying the evolution of Alfvén waves and firehose-like instabilities are the two plausible generation mechanisms of switchbacks. 

In contrast, many pieces of evidence suggest that the switchbacks originate from the Sun. Based on the estimations of the size and the orientations of magnetic field deflections of switchbacks observed by PSP, \citet{horbury-2020} indicate that these events could be the manifestations of coronal jets that are released via near-Sun impulsive interchange reconnection processes, which occur between open and closed magnetic field lines \citep{fisk-2001, yamauchi-2004, fisk-2005}. This is backed by previous coronal jet observations, which suggest that the jets could erupt and propagate to the PSP space and appear as switchbacks \citep{nistico-2010, sterling-2020, dePablos-2022}. The hypothesis that switchbacks are formed by interchange reconnection process near the Sun is supported by several other works \citep[e.g.][]{fisk-2020, zank-2020, liang-2021, fargette-2022}. The reduction of switchbacks in the sub-Alfvénic solar wind, which is first reported as the PSP entered the solar corona on 2021 April 28, may also correlate with lower magnetic reconnection rates on the surface of the Sun \citep{kasper-2021}. Furthermore, \citet{telloni-2022} report possible direct evidence of switchback in the solar corona by using the Metis coronagraph onboard Solar Orbiter, and they suggest that this switchback is favorably produced by interchange reconnection process. 
In addition, the widespread kink/Alfvén waves in the corona \citep{tomczyk-2007, tian-2012} provides another possibility that the Alfvénic switchbacks could be highly kinked Alfvén waves that generated in the corona and survived out to PSP distances, as shown by the latest simulations on Alfvénic fluctuations \citep{he-2020, tenerani-2020, shoda-2021}. Moreover, the characteristics of the cross helicity and the directions of electron heat flux in switchbacks also support the idea that the switchbacks could be local folds of magnetic field lines that originate from the corona and travel past the spacecraft \citep{mcmanus-2020}. Besides, some works further imply the possible relationship between the patches of switchbacks and the solar supergranulation and granulation \citep{bale-2021, fargette-2021, fargette-2022, shi-2022}. 

The compositional properties of switchbacks are pivotal to tracing their source regions \citep{kasper-2007, abbo-2016, huang-2016a, huang-2018}, but the associated works are still limited due to data calibrations during early encounters. \cite{bale-2021} find that the alpha to proton abundance ratio increases in several patches of switchbacks, and they thus propose an origin of switchbacks from above the transition region via interchange reconnection by combining other signatures. \citet{mcmanus-2022} further find there is no consistent compositional signature difference inside the switchbacks versus outside them by analyzing 92 switchbacks identified from PSP encounters 3 and 4. However, with the alpha data available in current encounters, we may find new clues to the origins of switchbacks based on the statistical study of the switchbacks in different solar wind conditions.

The stability of switchback structures associates directly with the evolution and the origin of switchbacks. 
The switchbacks are "S-shaped" Alfvénic structures \citep{kasper-2019}, but whether their structures are stable has not been understood. 
The switchbacks are not rare in previous observations by Helios at and beyond 0.3 au \citep{horbury-2018, woolley-2020} and other spacecraft near 1 au \citep[e.g.][]{kahler-1996, mccomas-1996, crooker-2004SB, huang-2017}. The striking difference of PSP observations is the unexpected prevalence of switchbacks in the near-Sun environment \citet{kasper-2019}, which implies that many switchbacks should disappear from PSP space to beyond. In regard to this fact, there are several questions that need to be answered. If the switchbacks originate from the Sun, then we need to explain how they could propagate to the PSP space but disappear when traveling farther. If the switchbacks are formed locally, we need to know the reason why they are formed locally and why they are more prevalent in the inner heliosphere.
In general, the folded magnetic field structures will stretch to radial as the solar wind propagates in the interplanetary space, thus there must be some mechanisms to help avoid the relaxations of the S-shaped structures of switchbacks. \citet{rasca-2021} predict the switchbacks may survive until around 60 $\mathrm{R_S}$ based on the evolutions of the magnetic field and speed changes in the boundaries of switchbacks. Some studies indicate that current sheets exist in the switchback boundaries \citep{farrell-2020, krasnoselskikh-2020, martinovic-2021}, which implies the presence of strong electric fields in the layers to keep the system quasi-stable. These results imply that the boundaries of switchbacks may play a role to prevent the switchbacks from being destroyed. Consequently, it is valuable to investigate the structures of switchbacks to understand their stability. 

In this work, we will focus on the structures and alpha particle variations in switchbacks. Following the method of \citet{kasper-2019}, we identify thousands of switchbacks with PSP observations during encounters 1-8 (E1-E8)\footnote{E3 is excluded due to data gaps.}$^{,}$\footnote{The switchback event lists are released to the SWEAP data repository: \url{http://sprg.ssl.berkeley.edu/data/psp/data/sci/sweap/lists/Switchbacks_PSP/}.}. Based on the analysis of the current sheet distributions and alpha particle characteristics, we investigate the structure and origin of the switchbacks. The data are described in Section \ref{sec:data}. Section \ref{sec:results} includes the method to identify switchbacks, the distributions of small-scale current sheets and also the alpha signatures of switchbacks in E4-E8. We present the discussion and summary in Section \ref{sec:disc} and Section \ref{sec:sum}, respectively.

\section{Data} \label{sec:data}
The Solar Wind Electrons, Alphas, and Protons (SWEAP) instrument suite \citep{kasper-2016} and the FIELDS instrument suite \citep{bale-2016} onboard PSP provide the data used in this work. SWEAP includes the Solar Probe Cup (SPC) \citep{Case-2020}, Solar Probe Analyzer for Electrons (SPAN-E) \citep{whittlesey-2020}, and Solar Probe Analyzer for Ions (SPAN-I) \citep{livi-2022}. SWEAP is designed to measure the velocity distributions of solar wind electrons, protons, and alpha particles \citep{kasper-2016}. FIELDS is designed to measure DC and fluctuation magnetic and electric fields, plasma wave spectra and polarization properties, the spacecraft floating potential, and solar radio emissions \citep{bale-2016}.

In this work, we use the magnetic field data from E1 to E8 to study the distributions of small-scale current sheets in switchbacks, and we use the fitted proton and alpha data from SPAN-I that are available since E4 to investigate the alpha characteristics in switchbacks. SPAN-I measures three-dimensional velocity distribution functions of the ambient ion populations in the energy range from several eV $q^{-1}$ to 20 keV $q^{-1}$ at a maximum cadence of 0.437 s, and it has a time of flight section that enables it to differentiate the ion species \citep{kasper-2016}. The details of the fitted proton and alpha data are described in \citet{finley-2020}, \citet{livi-2022}, and \citet{mcmanus-2022}. However, the SPAN-I measurements used here are from low-cadence downlinked data, and the time resolutions of the fitted proton and alpha data are 6.99 s and 13.98 s, respectively \citep{finley-2020, verniero-2020}. The FIELDS instrument collects high-resolution vector magnetic fields with variable time resolutions. The 4 samples per cycle (i.e. 4 samples per 0.874 s) data are used here.

\section{Results \label{sec:results}}

\subsection{Switchback identification \label{sec:swbid}}

\begin{figure}
\epsscale{1.}
\plotone{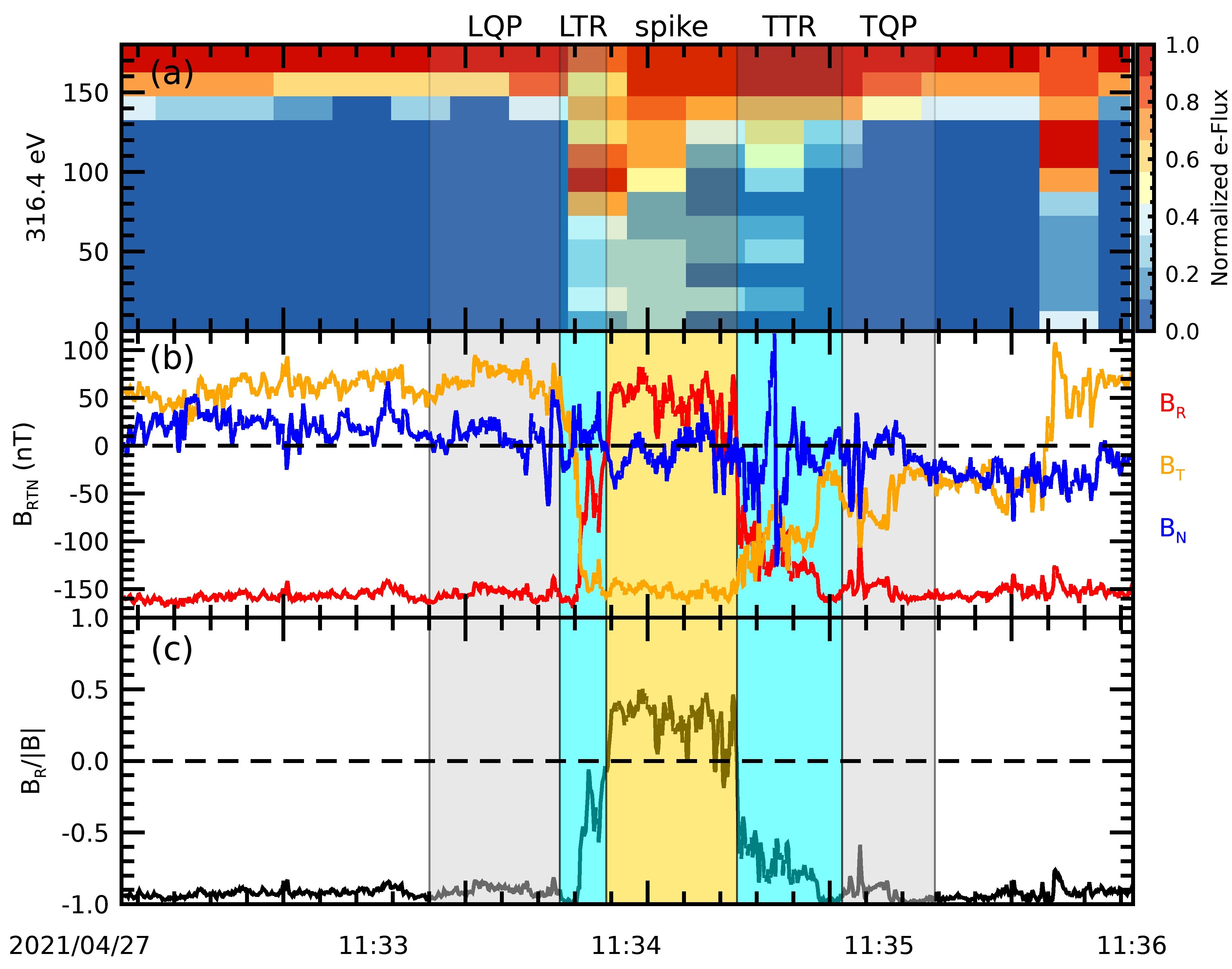}
\caption{An example of the switchback. From top to bottom, the panels show the normalized pitch angle distributions of suprathermal electrons (e-PADs) at an energy of 316.4 eV, the magnetic field components in RTN coordinates, and the radial to total magnetic field ratio, respectively. From left to right, the five shaded regions mark leading quiet period (LQP), leading transition region (LTR), spike, trailing transition region (TTR), and trailing quiet period (TQP), respectively. }
\label{fig:swbexp}
\end{figure}

\begin{figure}
\epsscale{1.2}
\plotone{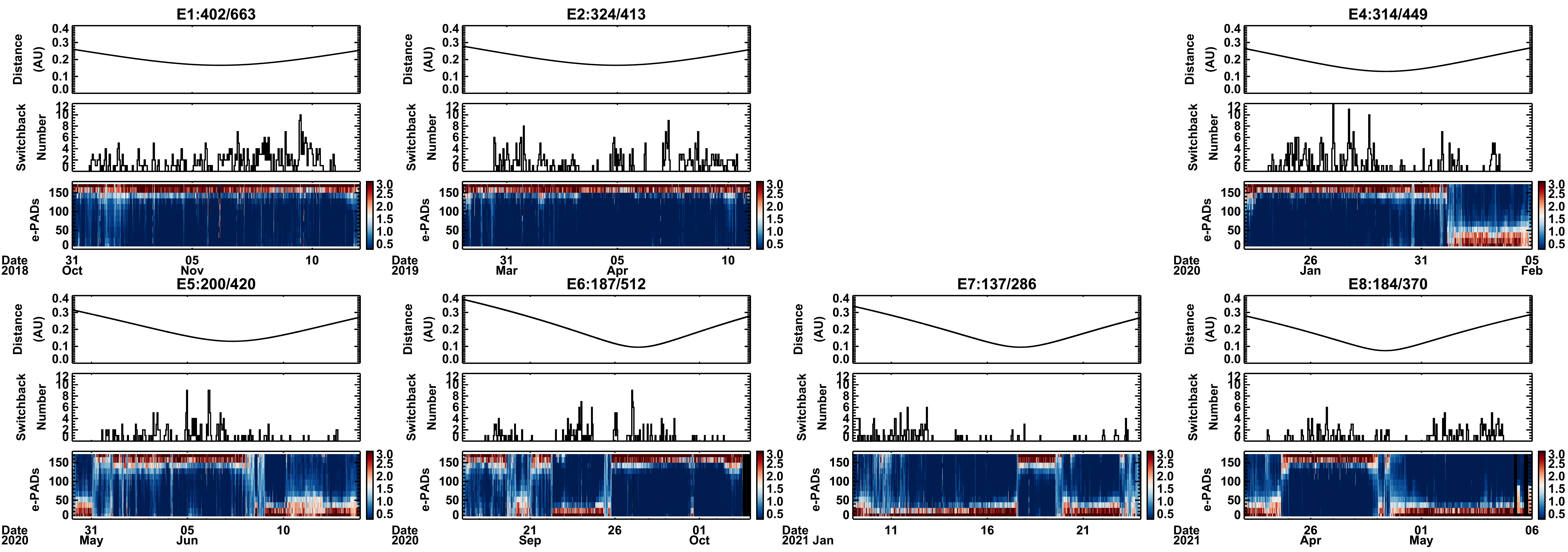}
\caption{Overview of the switchbacks identified in each encounter. In each figure, the top panel shows the distance of the spacecraft to the Sun, the middle panel shows the switchback number in each 1 hr interval, and the bottom panel shows the normalized electron pitch angle distributions (e-PADs). In the title of each panel, the number of good switchbacks vs. the total number of candidates identified in each encounter is given.}
\label{fig:swbdist}
\end{figure}

Following the method of \citet{kasper-2019}, we separate a switchback into five parts: leading/trailing quiet period (LQP/TQP), leading/trailing transition region (LTR/TTR), and spike (i.e. the core structure of switchback). Figure \ref{fig:swbexp} shows an example of the switchback observed on 2021 April 27 during E8. Panel (a) presents the pitch angle distributions of suprathermal electrons (e-PADs) at an energy of 316.4 eV. Panel (b) displays the magnetic field components in RTN coordinates, and panel (c) indicates the variations of the radial magnetic field component ($B_R$) to the total magnetic field strength ($|B|$), i.e. $B_R/|B|$. The spike is generally characterized by a fully magnetic field reversal, as shown by the yellow-shaded region, where the $B_R/|B|$ changes polarity while the dominant e-PADs keep the same. The two cyan-shaded regions represent the transition regions, and they indicate the magnetic field rotates from the quiet period to the spike and usually contains large-amplitude fluctuations. The LTR (TTR) locates between the spike and the LQP (TQP), which is the steady ambient solar wind as indicated by the grey-shaded regions. 

We use a two-step method to identify the switchbacks in each encounter. In the first step, we develop an automatic algorithm to search the candidates of switchbacks in about 11 days around the perihelion. The search interval is adjusted to be consistent with the time period when the SWEAP provides high time-resolution plasma observations. In the beginning, we need to find the heliospheric current sheet crossings, which mark the boundaries of sectors that have different polarities \citep{crooker-2004HPS, lavraud-2020, szabo-2019}. Then, we search the candidates for switchbacks in the sectors that have the same polarity. In order to find prominent reversals and also as many candidates as we can, we require the $B_R/|B|$ to change significantly from the quiet solar wind to the spike: (1) when the polarity of the sector is negative, we require $B_R/|B|>-0.25$ in the spike and $B_R/|B|<-0.85$ in the quiet period; (2) when the sector polarity is positive, we need $B_R/|B|<0.25$ in the spike and $B_R/|B|>0.85$ in the quiet period. We further demand there are at least three data points in the spike region. In the second step, we manually select the better switchbacks from the candidates. In this part, we use two criteria to select the better ones. On one hand, we require the e-PADs not to change their main distributions, which can help to verify the reversal of magnetic field lines is real rather than current sheets. On the other hand, we also compare the magnetic field components with the velocity components to find relatively high Alfvénic switchbacks. Consequently, we generate the lists of switchbacks for each encounter. We note that the switchbacks we identified in E1 and E2 have been used in studies by \citet{martinovic-2021} and \citet{akhavan-2021}, and our switchback lists from E1 to E8 have supported the work by \citet{akhavan-2022}. 

As a result, Figure \ref{fig:swbdist} shows an overview of the switchbacks in each encounter, with the panels from top to bottom showing the distance of the spacecraft to the Sun, the switchback number in each 1 hr interval, and the e-PADs. For the seven encounters, we find 663 switchback candidates in E1, 413 in E2, 449 in E4, 420 in E5, 512 in E6, 286 in E7, and 370 in E8. Among them, the good events selected for study are 402 in E1, 324 in E2, 314 in E4, 200 in E5, 187 in E6, 137 in E7, and 184 in E8. This means 1748 out of 3113 candidates are good switchbacks that suit further study. From the middle panel in each figure, we can also see the signature that the switchbacks distribute as patches \citep{horbury-2020, woolley-2020}. We also note that the switchbacks are profoundly reduced near the heliospheric current sheet crossings (where e-PADs change directions) in E4 to E8, one reason is that the multiple current sheet crossings \citep{szabo-2019} make it harder to identify the switchbacks.

\subsection{Current sheets in switchbacks \label{sec:subswb}}

\begin{figure}
\epsscale{1.2}
\plotone{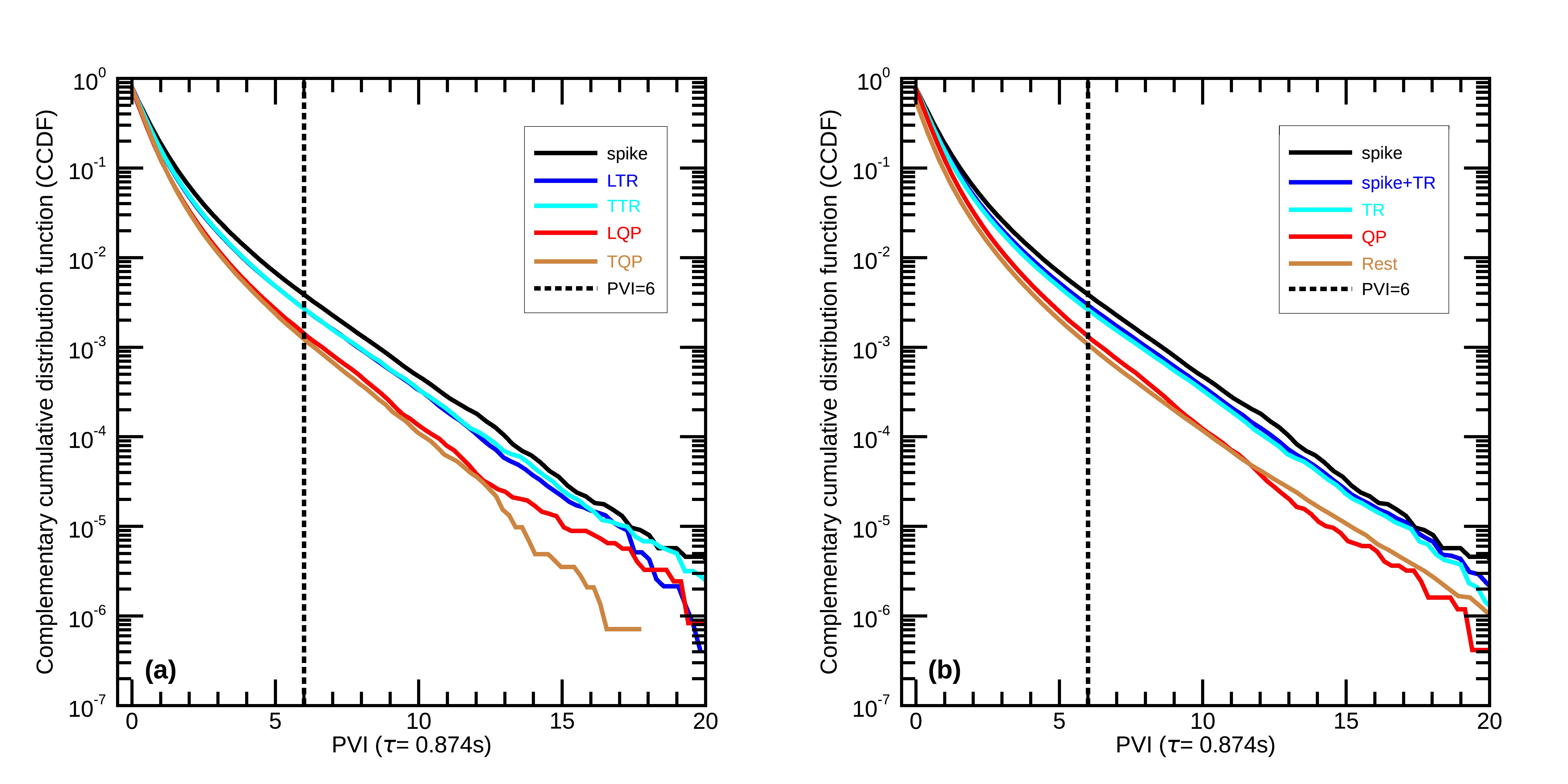}
\caption{The complementary cumulative distribution function (CCDF) of PVI values in different parts of switchbacks. In panel (a), spike, leading transition region (LTR), trailing transition region (TTR), leading quiet period (LQP), and trailing quiet period (TQP) are represented by black, blue, cyan, red, and brown colors, respectively. In panel (b), spike, combined region of spike and transition regions (spike+TR), combined LTR and TTR (TR), combined LQP and TQP (QP), and rest solar wind are represented by black, blue, cyan, red and brown colors, respectively. The vertical dashed lines in both panels mark the PVI value equal to 6. The switchbacks are from encounters E1-E8. }
\label{fig:PVIall}
\end{figure}

There are different methods to find current sheets in the solar wind, here we use the partial variance of increments (PVI) method, which is a reliable tool to identify such structures \citep[e.g.][]{greco-2008, podesta-2017, greco-2018, chhiber-2020, qudsi-2020}. The PVI at time \textit{s} is defined as the magnitude of the change in the magnetic field vector $\textbf{B}$ over a time lag $\tau$, thus 

\begin{equation}
\label{eq:pvieq}
 \mathrm{PVI_{s, \tau}} = \frac{|\Delta \textbf{B}(s, \tau)|}{\sqrt{\langle |\Delta \textbf{B}(s, \tau)|^2 \rangle}}  ,
\end{equation}

where $\Delta \textbf{B}(s, \tau) = \textbf{B}(s + \tau) - \textbf{B}(s)$ is the increment and $\langle ... \rangle$ represents an average over a suitable time period. In this work, we follow \citet{qudsi-2020} to set the time lag $\tau$ and the average interval as 0.874 s and 8 hrs (several times of the correlation time \citep{chen-2020}), respectively. 

According to previous studies, $\mathrm{PVI} > 3$ indicates non-Gaussian structures, $\mathrm{PVI} > 6$ means current sheets, and $\mathrm{PVI} > 8$ suggests reconnection sites \citep[][and references therein]{greco-2018, chhiber-2020}. Therefore, we choose PVI values that are larger than 6 to identify current sheets. Due to the high time resolution of magnetic field data and the cross trajectory of spacecraft, several adjacent data points with $\mathrm{PVI} > 6$ could belong to the same current sheet. We can either use the data points or the cases of current sheets to study the distributions. Here, we identify a current sheet case if the data point with $\mathrm{PVI} > 6$ is the maximum value in about 10 s, and the criterion is set due to only about 1.7\% of the selected switchbacks having a waiting time shorter than 10 s. We also find the results are similar between the two methods, here we only present the distributions of the cases of current sheets in the switchbacks. For the seven encounters, the corresponding current sheet cases therein are 2461, 2438, 2541, 2834, 3388, 3670, and 3038, respectively.  

Figure \ref{fig:PVIall} shows the complementary cumulative distribution function (CCDF, i.e. the probability of a variable that is above a threshold) of the PVI values in different parts of switchbacks. Panel (a) shows the CCDF of PVI values in the five parts of switchbacks (spike, LTR/TTR, LQP/TQP), represented by the different color lines as shown by the legend. Panel (b) exhibits the CCDF in combined regions of switchbacks by ignoring the asymmetries in leading and trailing edges, and we also include the rest of solar wind outside switchbacks ('Rest' in the figure) in the 11 days during each encounter for a comparison. TR means the combined LTR and TTR, and QP means the combined LQP and TQP. A similar figure to indicate the PVI variations in E1 and E2 are included in \citet{martinovic-2021}. Panel (a) denotes that large PVI values are more prevalent in spike (black line) and transition regions (blue and cyan lines) than that in quiet periods (red and brown lines), which is reasonable because the magnetic field generally changes smoothly in quiet periods. Moreover, it seems the leading and trailing locations of both quiet periods and transition regions do not produce large asymmetric distributions of current sheets ($\mathrm{PVI} > 6$). Panel (b) shows that the rest solar wind (brown line) has comparable current sheets with QP (red line). In addition, the spike (black line) has more current sheet densities than TR (cyan line), but the current sheet densities in the regions that include major magnetic field rotations (spike+TR, blue line) are much closer to the TR. This is a result of the longer duration but lower current sheet densities of TR, which may imply that transition regions play an important role in maintaining the structures of switchbacks. Overall, the distributions of current sheets in combined regions are consistent with the results present in panel (a). 

Table \ref{tab:PVIdis} lists the distributions of current sheets in different regions of switchbacks in different encounters. During each encounter, the numbers of current sheets ($N_{cs}$) are displayed in the first row. We then normalize the duration of each part of switchbacks to the total duration of spikes to get their relative durations ($D_{spike}$) and calculate the density of current sheets ($\rho_{cs}$), which is defined as the ratio of the case number to the relative duration ($\rho_{cs} = N_{cs} / D_{spike}$). The relative densities in each encounter are further described as percentages ($R$) in comparison with the current sheet density in spike for better visualization, whereas the $\hat{\textbf{R}}$ represents the relative current sheet density normalized by that in the spike of E1 for further comparison among different encounters. The results of all encounters are shown at the bottom of the table. 
This table reveals several features:

\begin{enumerate}
\item The current sheets are most prevalent in spikes. During each encounter, we can see the density $R$ is largest in spike and smallest in quiet periods, whereas the transition regions have intermediate density. This is consistent with the results shown in Figure \ref{fig:PVIall}. In all encounters, $R$ is 65.9\% in TR and 33.1\% in QP as compared with that in the spike. 

\item The current sheet distributions in leading and trailing boundaries show slight asymmetry. In each encounter, we can see the current sheet densities in either the quiet period or transition region are similar in leading and trailing edges, but $R$ is slightly different on both sides. For all encounters, the $R$ is 35.2\% versus 29.7\% in LQP and TQP, whereas 62.1\% versus 69.9\% in LTR and TTR, respectively. The LTR has larger $R$ than TTR during E1 and E2, but the circumstance reverses in later encounters except for E6, when the LTR and TTR have nearly the same $R$. The perihelia of E1 and E2 are $\sim$0.17 au, while the perihelia are much closer to the Sun in later encounters. The likely systematic changes imply the transition regions may evolve with radial distances from the Sun, which still needs future observations. 

\item The current sheet densities vary in different encounters. Focusing on the normalized current sheet densities $\hat{\textbf{R}}$, we can see the current sheet densities change significantly in different encounters. The $\hat{\textbf{R}}$ in spike are above 1 in E2-E5 but below 1 in E6-E8, while the transition regions show much more variable $\hat{\textbf{R}}$ in different encounters. However, in comparison with spike and transition regions, the quiet periods and rest solar wind have the least intensive current sheets but more uniform distributions between different encounters. As the PSP dives deeper toward the Sun, the current sheet densities increase remarkably in both quiet periods and rest solar wind, implying the switchbacks may stay in a more pristine state when closing to the Sun.   

\item The density in spike+TR region is determined by the TR. From the bottom of this table, we can see the current sheet density $R$ in TR is 65.9\% of that in spike, resulting in a 75.7\% level in spike+TR. This is a natural mathematical consequence of the longer duration but lower $R$ in TR. But this result implies that transition regions may play a more important role in maintaining the structure of switchbacks than spike. 

\item The rest of solar wind has slightly larger current sheet densities than that in the quiet periods. The QP and the rest solar wind have comparable current sheets densities, which is reasonable due to the QP solar wind is much closer to the ambient solar wind. Specifically, the rest of solar wind has higher $R$ than QP in all encounters except E4. Additionally, the current sheet densities in both QP and rest solar wind show a general decreasing trend with distance from the Sun, but the current sheets in QP seem to vanish more rapid than that in the rest solar wind. The results suggest that the switchback structures may dissipate starting from the QP as they propagate outward. 
\end{enumerate}


\begin{deluxetable}{|c|c|c|c|c|c|c|c|c|c|c|}
\tablecaption{The distributions of current sheets in different regions of switchbacks. \label{tab:PVIdis}}
\tablecolumns{45}
\tablenum{1}
\tablewidth{750pt}
\tablehead{
\multicolumn{1}{|c}{} &       & \multicolumn{1}{|c}{spike}   & 
\multicolumn{1}{|c}{LQP}      & \multicolumn{1}{|c}{LTR} & 
\multicolumn{1}{|c}{TTR}      & \multicolumn{1}{|c|}{TQP}  &
\multicolumn{1}{|c}{spike+TR} & \multicolumn{1}{|c|}{TR}   &
\multicolumn{1}{|c}{QP}       & \multicolumn{1}{|c|}{Rest} 
}
\startdata
 E1 &  $N_{cs}$      &   375		 & 24		&  355		 &  241		    &  70		&  939		&  575		&  94		&  1239      \\
    &  $D_{spike}$   & 1 (61108.3 s)	 & 0.437	&  1.508	 &  1.101	    &  0.878	&  3.414	&  2.500	&  1.264	&  9.969     \\
    &  $\rho_{cs}$   &   375.000	 & 54.920	&  235.411	 &  218.892		&  79.727	&  275.044	&  230.000	&  74.367	&  124.285   \\
    &  $R$       	&   \textbf{1.000}	 & \textbf{0.146}	&  \textbf{0.628}	 &  \textbf{0.584}		&  \textbf{0.213}	& \textbf{0.733}		&  \textbf{0.613}	&  \textbf{0.198}	&  \textbf{0.331}     \\
    &  $\hat{\textbf{R}}$   	&   \textbf{1.000}	 & \textbf{0.146}	&  \textbf{0.628}	 &  \textbf{0.584}		&  \textbf{0.213}	& \textbf{0.733}		&  \textbf{0.613}	&  \textbf{0.198}	&  \textbf{0.331}      \\
\hline              
 E2 &  $N_{cs}$      &   391		 & 29		&  223		 &  147		    &  51		&  733		&  362		&  74		&  1468      \\
    &  $D_{spike}$   &  1 (58073.8 s)	 & 0.351	&  1.100	 &  0.836	    &  0.661	&  2.861	&  1.912	&  0.936	&  12.780    \\
    &  $\rho_{cs}$ 	&   391.000	 & 82.513	&  202.761	 &  175.882	    &  77.190	&  256.179	&  189.303	&  79.026	&  114.865   \\
    &  $R$          &   \textbf{1.000}	 & \textbf{0.211}	&  \textbf{0.519}	 &  \textbf{0.450}	    &  \textbf{0.197}	&  \textbf{0.655}	&  \textbf{0.484}	&  \textbf{0.202}	&  \textbf{0.294}     \\
    &  $\hat{\textbf{R}}$   	&   \textbf{1.097}	 & \textbf{0.232}	&  \textbf{0.569}	 &  \textbf{0.493}		&  \textbf{0.216}	&  \textbf{0.719}	&  \textbf{0.531}	&  \textbf{0.222}	&  \textbf{0.322}    \\
\hline                
 E4 &  $N_{cs}$      &   227		 & 79		&  225		 &  318		    &  72		&  742		&  530		&  144		&  1486      \\
    &  $D_{spike}$   &  1 (34967.0 s)	 & 0.857	&  1.489	 &  1.503	    &  0.862	&  3.841	&  2.910	&  1.674	&  20.467    \\
    &  $\rho_{cs}$   &   227.000	 & 92.223	&  151.083	 &  211.636	    &  83.536	&  193.202	&  182.142	&  86.013	&  72.606    \\
    &  $R$          &   \textbf{1.000}	 & \textbf{0.406}	&  \textbf{0.666}	 &  \textbf{0.932}	    &  \textbf{0.368}	&  \textbf{0.851}	&  \textbf{0.802}	&  \textbf{0.379}	&  \textbf{0.320}     \\
    &  $\hat{\textbf{R}}$   	&   \textbf{1.058}	 & \textbf{0.430}	&  \textbf{0.704}	 &  \textbf{0.986}		&  \textbf{0.389}	&  \textbf{0.900}	&  \textbf{0.849}	&  \textbf{0.401}	&  \textbf{0.338}    \\
\hline                
 E5 &  $N_{cs}$      &   279		 & 77		&  234		 &  330		    &  74		&  748		&  529		&  147		&  1862      \\
    &  $D_{spike}$   &   1 (36541.9 s) & 1.047	&  1.839	 &  1.943	    &  1.028	&  4.388	&  3.583	&  2.000	&  23.116    \\
    &  $\rho_{cs}$   &   279.000	 & 73.563	&  127.235	 &  169.831	    &  71.965	&  170.469	&  147.647	&  73.502	&  80.551    \\
    &  $R$          &   \textbf{1.000}	 & \textbf{0.264}	&  \textbf{0.456}	 &  \textbf{0.609}	    &  \textbf{0.258}	&  \textbf{0.611}	&  \textbf{0.529}	&  \textbf{0.263}	&  \textbf{0.289 }    \\
    &  $\hat{\textbf{R}}$   	&   \textbf{1.244}	 & \textbf{0.328}	&  \textbf{0.567}	 &  \textbf{0.757}	 	&  \textbf{0.321}	&  \textbf{0.760}	&  \textbf{0.658}	&  \textbf{0.328}	&  \textbf{0.359}     \\
\hline                
 E6 &  $N_{cs}$      &   312		 & 143		&  319		 &  325		    &  91		&  864		&  605		&  225		&  2276      \\
    &  $D_{spike}$   & 1 (65198.9 s)  & 0.851	&  1.347	 &  1.387	    &  0.797	&  3.370	&  2.540	&  1.504	&  14.313    \\
    &  $\rho_{cs}$   &   312.000	 & 168.112	&  236.742	 &  234.281	    &  114.154	&  256.356	&  238.175	&  149.634	&  159.017   \\
    &  $R$          &   \textbf{1.000}	 & \textbf{0.539}	&  \textbf{0.759}	 &  \textbf{0.751}	    &  \textbf{0.366}	&  \textbf{0.822}	&  \textbf{0.763}	&  \textbf{0.480}	&  \textbf{0.510}     \\
    &  $\hat{\textbf{R}}$   	&   \textbf{0.780}	 &  \textbf{0.420}	&  \textbf{0.592}	 &  \textbf{0.586}		&  \textbf{0.285}	&  \textbf{0.641}	&  \textbf{0.595}	&  \textbf{0.374}	&  \textbf{0.397}     \\
\hline                
 E7 &  $N_{cs}$      &   384		 & 95		&  288		 &  384		    &  104		&  972		&  632		&  196		&  2499      \\
    &  $D_{spike}$   &  1 (64974.3 s)	 & 0.802	&  1.217	 &  1.346	    &  0.840	&  3.205	&  2.402	&  1.511	&  14.361    \\
    &  $\rho_{cs}$ 	&   384.000	 & 118.391	&  236.578	 &  285.198	    &  123.802	&  303.282	&  263.106	&  129.677	&  174.016   \\
    &  $R$          &   \textbf{1.000}	 & \textbf{0.308}	&  \textbf{0.616}	 &  \textbf{0.743}	    &  \textbf{0.322}	&  \textbf{0.790}	&  \textbf{0.685}	&  \textbf{0.338}	&  \textbf{0.453 }    \\
    &  $\hat{\textbf{R}}$   	&   \textbf{0.963}	 & \textbf{0.297}	&  \textbf{0.594}	 &  \textbf{0.716}		&  \textbf{0.311}	&  \textbf{0.761}	&  \textbf{0.660}	&  \textbf{0.325}	&  \textbf{0.436 }    \\
\hline                
 E8 &  $N_{cs}$      &   332		 & 122		&  252		 &  282  &  95		&  735		&  470		&  198		&  2074      \\
    &  $D_{spike}$   &  1 (62161.1 s)  & 0.741	&  1.070	 &  1.056	    &  0.744	&  2.629	&  1.888	&  1.329	&  11.314    \\
    &  $\rho_{cs}$   &   332.000	 & 164.567	&  235.548	 &  267.057	    &  127.696	&  279.521	&  248.901	&  148.983	&  183.316   \\
    &  $R$         	&  \textbf{1.000}	 & \textbf{0.496}	&  \textbf{0.709}	 &  \textbf{0.804}	    &  \textbf{0.385}	&  \textbf{0.842}	&  \textbf{0.750}	&  \textbf{0.449}	&  \textbf{0.552}     \\
    &  $\hat{\textbf{R}}$   	&   \textbf{0.870} & \textbf{0.432}	&  \textbf{0.617}	 &  \textbf{0.700}		&  \textbf{0.335}	&  \textbf{0.733}	&  \textbf{0.653}	&  \textbf{0.391}	&  \textbf{0.481}     \\
\hline                
All &  $N_{cs}$      &   2300	 & 569		&  1896		 &  2027		&  557		&  5733		&  3703		&  1078		&  12904     \\
    &  $D_{spike}$   &  1 (383025.3 s) & 0.702	&  1.328	 &  1.261		&  0.816	&  3.292	&  2.443	&  1.415	&  14.311    \\
    &  $\rho_{cs}$ 	&   2300.000 & 810.235	&  1427.430	 &  1607.760	&  682.654	&  1741.610	&  1516.050	&  761.643	&  901.718   \\
    &  $R$          &  \textbf{1.000}	 & \textbf{0.352}	&  \textbf{0.621}	 &  \textbf{0.699}		&  \textbf{0.297}	&  \textbf{0.757}	&  \textbf{0.659}	&  \textbf{0.331}	&  \textbf{0.392}     \\
    &  $\hat{\textbf{R}}$   	&  \textbf{0.979}	 & \textbf{0.345}	&  \textbf{0.607}	 &  \textbf{0.684}		&  \textbf{0.290}	&  \textbf{0.741}	&  \textbf{0.645}	&  \textbf{0.324}	&  \textbf{0.384}     \\
\enddata
\tablenotetext{a}{$N_{cs}$ means the number of current sheets in each region of switchbacks. $D_{spike}$ is the relative duration of each region to the spike duration, and the spike duration is set to be 1 in each encounter with the actual cumulative duration listed.
$\rho_{cs} = N_{cs} / D_{spike}$ indicates the density of current sheets in each region of switchbacks. $R$ is the ratio of the current sheet density in each region to that in the spike in each encounter. $\hat{\textbf{R}}$ represents the relative current sheet density normalized by that in the spike of E1. }
\tablenotetext{b}{Spike: complete reversal of magnetic field polarity. LQP/TQP/QP: Leading/Trailing/Combined Quiet Period. LTR/TTR/TR: Leading/Trailing/Combined Transition Region. 
Rest: rest solar wind, which is the solar wind between the first and the last switchback by removing all switchbacks therein during each encounter. The current sheet number $N_{cs}$ in QP (TR) is generally smaller than the summed number in LQP and TQP (LTR and TTR) is caused by the overlap of some switchbacks. 
}  
\end{deluxetable}

\subsection{Helium signatures \label{sec:Helium}}

\begin{figure}
\epsscale{1.2}
\plotone{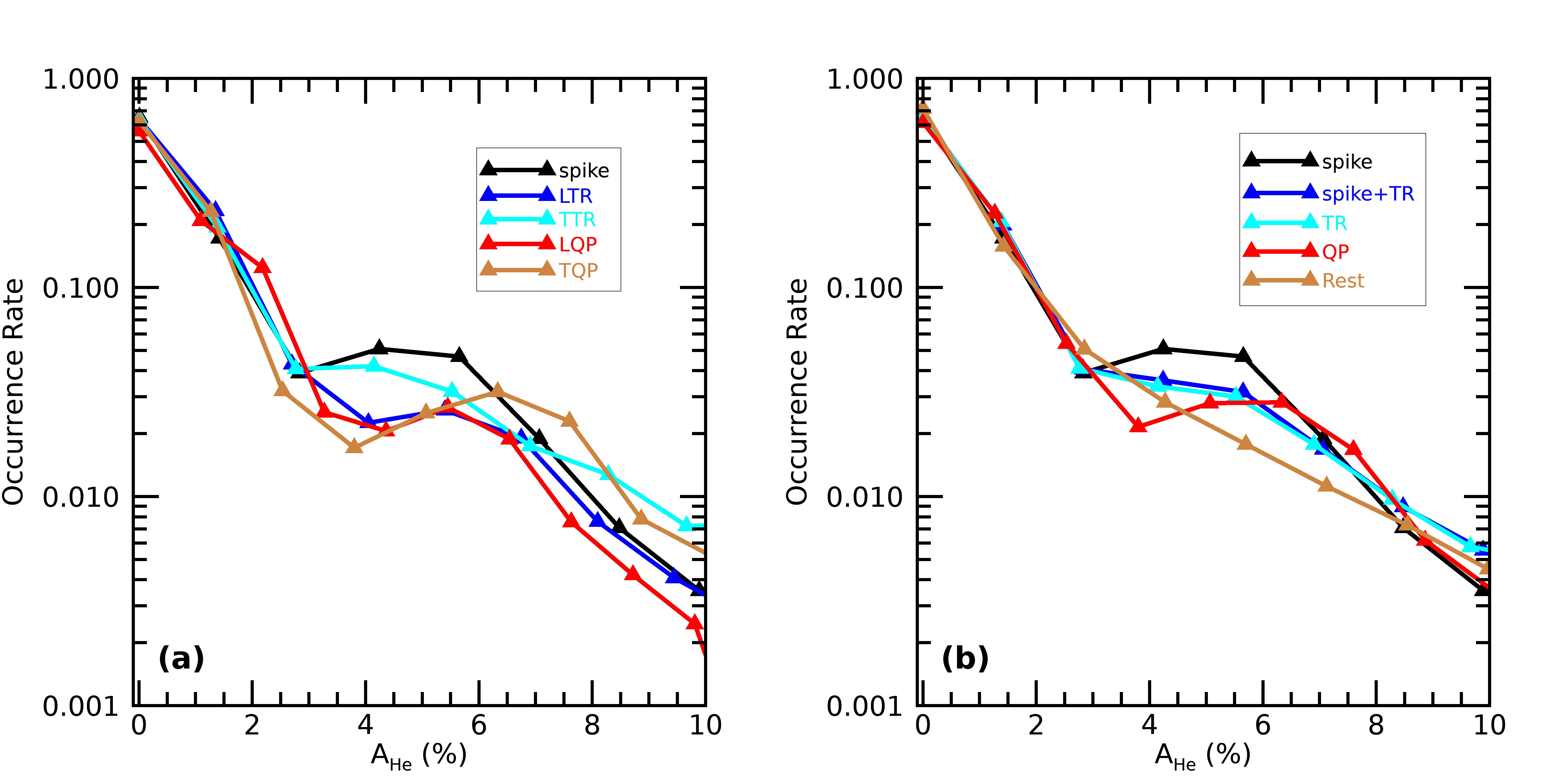}
\caption{The occurrence rates of alpha to proton abundance ratio ($\mathrm{A_{He}}$) inside and outside switchbacks. Panel (a) shows the $\mathrm{A_{He}}$ variations in different parts of switchbacks, and panel (b) compares their distributions in combined regions of switchbacks. Different regions of switchbacks are represented by different colors of lines, and the colors in panel (a) and panel (b) are the same to that in Figure \ref{fig:PVIall}, respectively. }
\label{fig:Nap}
\end{figure}

\begin{figure}
\epsscale{1.}
\plotone{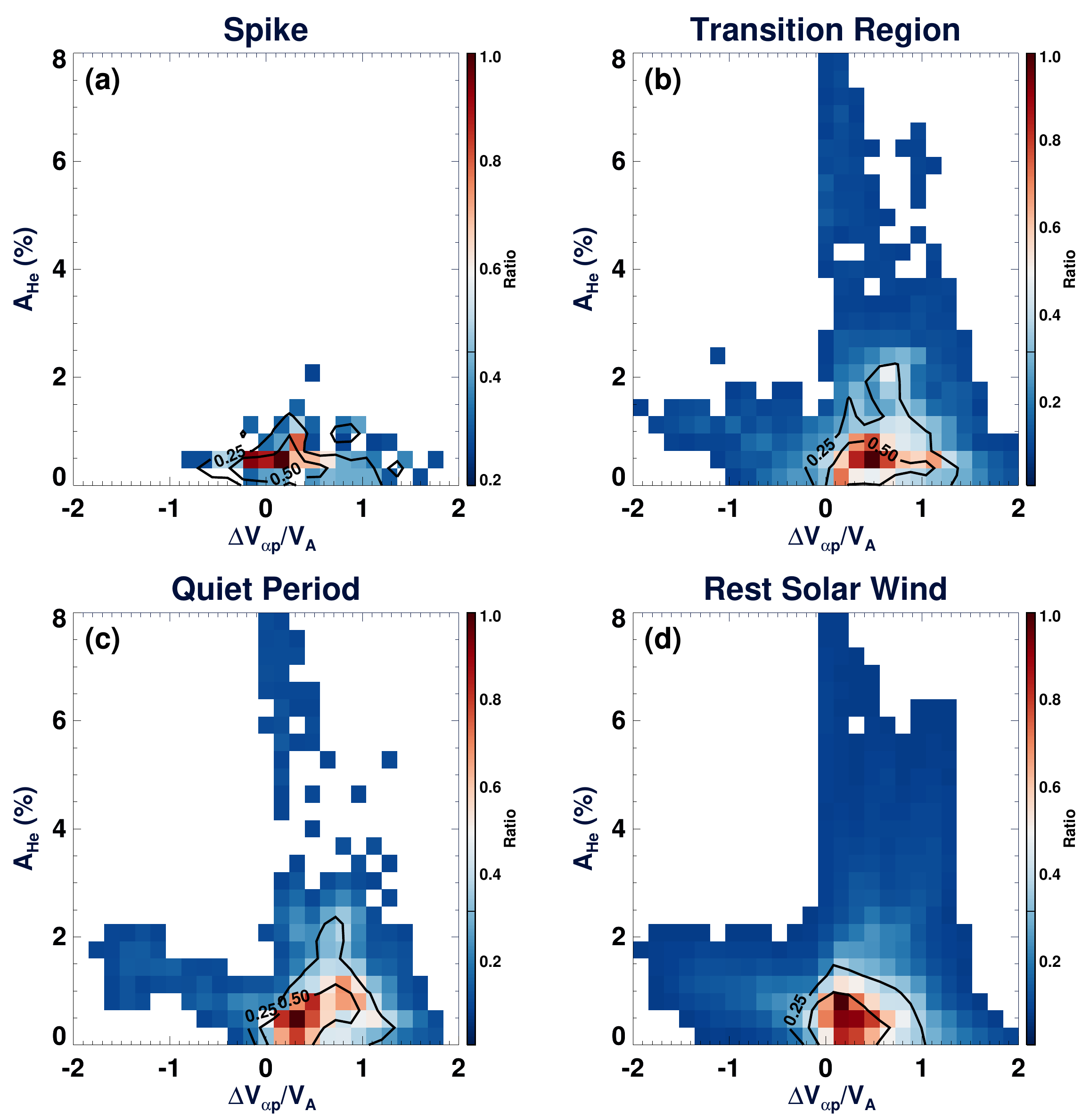}
\caption{The distributions of alpha to proton abundance ratios ($\mathrm{A_{He}}$) versus the alpha-proton differential speed normalized by local Alfvén speed ($\mathrm{\Delta V_{\alpha p}/V_A}$) inside and outside switchbacks. Panels (a) to (d) show the distributions in spike, combined transition region, combined quiet period, and rest solar wind, respectively. The colors in each panel indicate the ratio of data counts. The black lines indicate 25\% and 50\% measurement contours. }
\label{fig:NVap}
\end{figure}

The helium measurements give clues to the solar wind origins \citep[e.g.][]{aellig-2001, bochsler-2007, kasper-2012, huang-2016b, huang-2023SBSW, fu-2018}. The helium abundance ratio ($A_{He} = N_\alpha/N_p \times 100\%$) is usually enhanced in the fast solar wind and magnetic clouds, implying the helium-rich population originates from open magnetic field regions in the Sun \citep{borrini-1981, gosling-1981, suess-2009}. Nevertheless, the $\mathrm{A_{He}}$ in slow solar wind greatly correlates with solar activity, indicating that slow wind with decreased $\mathrm{A_{He}}$ (helium-poor population) at solar minimum should come from helmet streamer whereas slow wind with increased $\mathrm{A_{He}}$ at solar maximum mainly comes from active regions \citep{kasper-2007, kasper-2012, alterman-2018, alterman-2019}. 

Moreover, alpha-proton differential speed ($\mathrm{\Delta V_{\alpha p}}$) is large in the fast wind and comparable to local Alfvén speed ($\mathrm{V_A}$), but $\mathrm{\Delta V_{\alpha p}}$ is close to zero in slow solar wind \citep{marsch-1982, steinberg-1996, reisenfeld-2001, berger-2011}. Current studies indicate bimodal distributions of $\mathrm{A_{He}}$ versus $\mathrm{\Delta V_{\alpha p}/V_A}$ in the solar wind, with high $\mathrm{A_{He}}$ and high $\mathrm{\Delta V_{\alpha p}/V_A}$ population escaping directly along open magnetic field lines as described by wave-turbulence driven models, while low $\mathrm{A_{He}}$ and low $\mathrm{\Delta V_{\alpha p}/V_A}$ population releasing through magnetic reconnection processes \citep[][and references therein]{durovcova-2017, fu-2018, durovcova-2019}. Therefore, the characteristics of $\mathrm{A_{He}}$ and $\mathrm{\Delta V_{\alpha p}/V_A}$ in switchbacks could infer their possible origins. 

Following \citet{reisenfeld-2001} and \citet{fu-2018}, we define the $\mathrm{\Delta V_{\alpha p}}$ as the field-aligned differential speed, i.e. $\mathrm{\Delta V_{\alpha p}} = (v_{\alpha r}-v_{pr})/cos(\theta)$, where $v_{\alpha r}$ and $v_{pr}$ are the radial speeds of alpha particle and proton, respectively, and $\theta$ measures the angle of the magnetic field vector from the radial direction. Here, we further require $cos(\theta) = |B_r/B|$ to assure the derived differential speed is independent of magnetic field polarity, where $B_r$ and $B$ represent the radial component and the total strength of the magnetic field, respectively.
Moreover, the local Alfvén speed is calculated with $\mathrm{V_A}=|B|/\sqrt{\mu_0(N_pm_p+N_{\alpha}m_{\alpha})}$, where $\mu_0$ denotes the vacuum magnetic permeability, $N_p$ ($N_{\alpha}$) and $m_p$ ($m_{\alpha}$) are the number density and mass of proton (alpha particle), respectively. In the calculations, we use the electron density derived from the analysis of plasma quasi-thermal noise (QTN) spectrum measured by the FIELDS Radio Frequency Spectrometer \citep{pulupa-2017, moncuquet-2020} to replace $N_p$ by assuming charge neutrality and $\mathrm{A_{He}}$ is 4\%, which does not significantly change the $\mathrm{V_A}$ as $\mathrm{A_{He}}$ generally varies from 1\% to 8\% \citep{liu-2021, mostafavi-2022}. 

The fitted helium data are available since E4 and are of good quality due to the solar wind flows predominantly into the field of view of SPAN-I, thus we choose the alpha data from E4 to E8 for this part of work. The data quality is further verified by comparing the proton and alpha densities from SPAN-I with the QTN electron density. Figure \ref{fig:Nap} presents the $\mathrm{A_{He}}$ variations inside and outside switchbacks. Figure \ref{fig:Nap}(a) denotes the occurrence rates of $\mathrm{A_{He}}$ in the five regions of switchbacks, marked by the colors as shown in the legend. It shows that both helium-poor ($\mathrm{A_{He}} \sim 1.0\%$) and helium-rich ($\mathrm{A_{He}} \sim 6.0\%$) populations exist in different parts of switchbacks, suggesting the switchbacks may originate from both closed and open magnetic field regions in the Sun. This is also supported by previous studies that the switchbacks are found both in slow and fast solar winds \citep[e.g.][]{horbury-2018, horbury-2020, de-2020, bale-2021}. In addition, the $\mathrm{A_{He}}$ distributions show no significant difference inside and outside switchbacks, and the similar plasma indicates the switchback is an intact structure that has no big difference in different regions, which is consistent with latest work by \citet{mcmanus-2022}. Moreover, the helium-poor population overwhelms the helium-rich population in switchbacks, implying that solar wind from closed magnetic field regions contributes more to these switchbacks. The reason could be that the switchbacks are identified mainly from slow and intermediate speed solar wind during these encounters, as we find the solar wind with speed larger than 600 km/s, 500 km/s and 400 km/s is 0.31\%, 0.85\% and 6.71\% of the time, respectively. Furthermore, Figure \ref{fig:Nap}(b) displays the $\mathrm{A_{He}}$ variations in the combined regions of switchbacks. The $\mathrm{A_{He}}$ features are similar and more uniform in the combined regions as the asymmetry in leading and trailing edges are averaged. 

In Figure \ref{fig:NVap}, panels (a) to (d) show the variations of $\mathrm{A_{He}}$ versus the field-aligned $\mathrm{\Delta V_{\alpha p}/V_A}$ in spike, TR, QP and rest solar wind, respectively. The black lines in the figure indicate the 25\% and 50\% measurement contours. 
In the spike, the major solar wind shows low $\mathrm{A_{He}}$ ($<$ 2.0\%), whereas the $\mathrm{\Delta V_{\alpha p}/V_A}$ varies from -1 to 2. As introduced above, $\mathrm{\Delta V_{\alpha p}}$ is comparable to $\mathrm{V_A}$ in the fast wind but close to zero in the slow wind. \citet{mostafavi-2022} find that $\mathrm{\Delta V_{\alpha p}}$ increases toward the Sun but the magnitude is mainly below $\mathrm{V_A}$, and alpha particles usually move faster than protons near the Sun, based on PSP observations. However, the high $\mathrm{\Delta V_{\alpha p}/V_A}$ in the solar wind may associate with the very low local Alfvén speed when magnetic field lines change polarities, or it may infer the preferential acceleration of alpha particles \citep{isenberg-1983, kasper-2017}. We note that there are negative values of $\mathrm{\Delta V_{\alpha p}/V_A}$, which could be a result of waves that slow down alpha particles but accelerate the protons as the energy of alpha particles overtakes protons \citep{durovcova-2017}. Therefore, the low $\mathrm{A_{He}}$ and low $\mathrm{\Delta V_{\alpha p}/V_A}$ ($|\mathrm{\Delta V_{\alpha p}/V_A}|<0.7$) implies that most of the switchbacks could be formed by magnetic reconnection process, whereas the low $\mathrm{A_{He}}$ and high $\mathrm{\Delta V_{\alpha p}/V_A}$ ($|\mathrm{\Delta V_{\alpha p}/V_A}|>0.7$) indicates the possible preferential acceleration ($\mathrm{\Delta V_{\alpha p}}>0$) or slow down ($\mathrm{\Delta V_{\alpha p}}<0$) of alpha particles. 
The $\mathrm{A_{He}}$ - $\mathrm{\Delta V_{\alpha p}/V_A}$ distributions are similar in TR and QP. The dominant population shows low $\mathrm{A_{He}}$ ($< 4.0\%$) and low $\mathrm{\Delta V_{\alpha p}/V_A}$ ($|\mathrm{\Delta V_{\alpha p}/V_A}|<1.0$), but $\mathrm{\Delta V_{\alpha p}/V_A}$ varies in a large range from about -2 to 2. Additionally, the TR and QP solar winds also have a population with high $\mathrm{A_{He}}$ ($\sim$ 6.0\%) but low $\mathrm{\Delta V_{\alpha p}/V_A}$, which may suggest compressions in the transition regions that work to decrease the differential speed \citep{durovcova-2019}. \citet{krasnoselskikh-2020} and \citet{larosa-2021} also confirm the existence of compressible switchbacks. Therefore, in TR and QP, the magnetic reconnection mechanism works overwhelmingly to the wave-turbulence-driven mechanism, but compressions and/or waves are rich in these regions to change the $\mathrm{A_{He}}$ - $\mathrm{\Delta V_{\alpha p}/V_A}$ distributions. 
In the rest of solar wind, both $\mathrm{A_{He}}$ and $\mathrm{\Delta V_{\alpha p}/V_A}$ range from low to high values with predominant low $\mathrm{A_{He}}$ and low $\mathrm{\Delta V_{\alpha p}/V_A}$ population, indicating the reconnection mechanism dominates but other mechanisms also play a role.

\section{Discussion} \label{sec:disc}
\subsection{The stability of switchback structures \label{sec:structure}}

The switchbacks are much more abundant in the inner heliosphere than beyond. One reason could be that the PSP provides higher time-resolution measurements, which helps find more, especially, short-duration switchbacks. Another possible explanation for this is that the switchbacks may not be able to survive, if they are formed from the Sun, for a long heliocentric distance. If so, the stability of switchback structures could be a key to understanding the evolution of switchbacks. 

According to the distributions of small-scale current sheets in switchbacks, we could infer that these current sheets should contribute to stabilizing the switchback structures. In general, the small-scale current sheets are a measure of the braiding of the magnetic field within the switchbacks. The magnetic braiding was proposed to explain the coronal heating by \citet{parker-1972, parker-1988, parker-1994}, with the hypothesis that the random, complex motions of footpoints at the base of coronal loops due to the photospheric turbulence could twist and entangle the magnetic field, and the coronal heating happens as the braided magnetic fields inevitably relax to a force-free state \citep{schrijver-1998, wilmot-2015}. The relaxation of the braided field via magnetic reconnection processes would eventually result in a complex array of current sheets \citep{longbottom-1998, pontin-2011, ng-2012, rappazzo-2013}. \citet{prior-2016a, prior-2016b} construct different types of braided magnetic field and investigate their evolutions in background field, their detailed studies suggest that the braided internal structures can inhibit large-scale morphological changes. Therefore, the small-scale current sheets in switchbacks should also help to stabilize the switchback structures. 

Additionally, the S-shaped curvature of switchbacks may also help achieve stability. The S-shaped magnetic field could form two large-scale current sheets on both sides of a spike. Thus, the Lorentz force in the transverse direction could drag the magnetic structure together, and may further result in a magnetic component in this direction that works to stabilize the current sheets, which are well studied in different environments \citep[refer to the review papers][ and references therein]{zelenyi-2010, artemyev-2012, treumann-2013}. Furthermore, the compressions in transition regions, as we suggested in section \ref{sec:Helium}, could resist the magnetic tensions to prevent the S-shaped curvature from relaxing.

In conclusion, we suggest the small-scale current sheets in switchbacks and the S-shaped curvatures of switchbacks may work together to stabilize their structures and enable them to propagate stably into PSP space.

\subsection{The origins of switchbacks \label{sec:origin}}

\begin{figure}
\epsscale{1.0}
\plotone{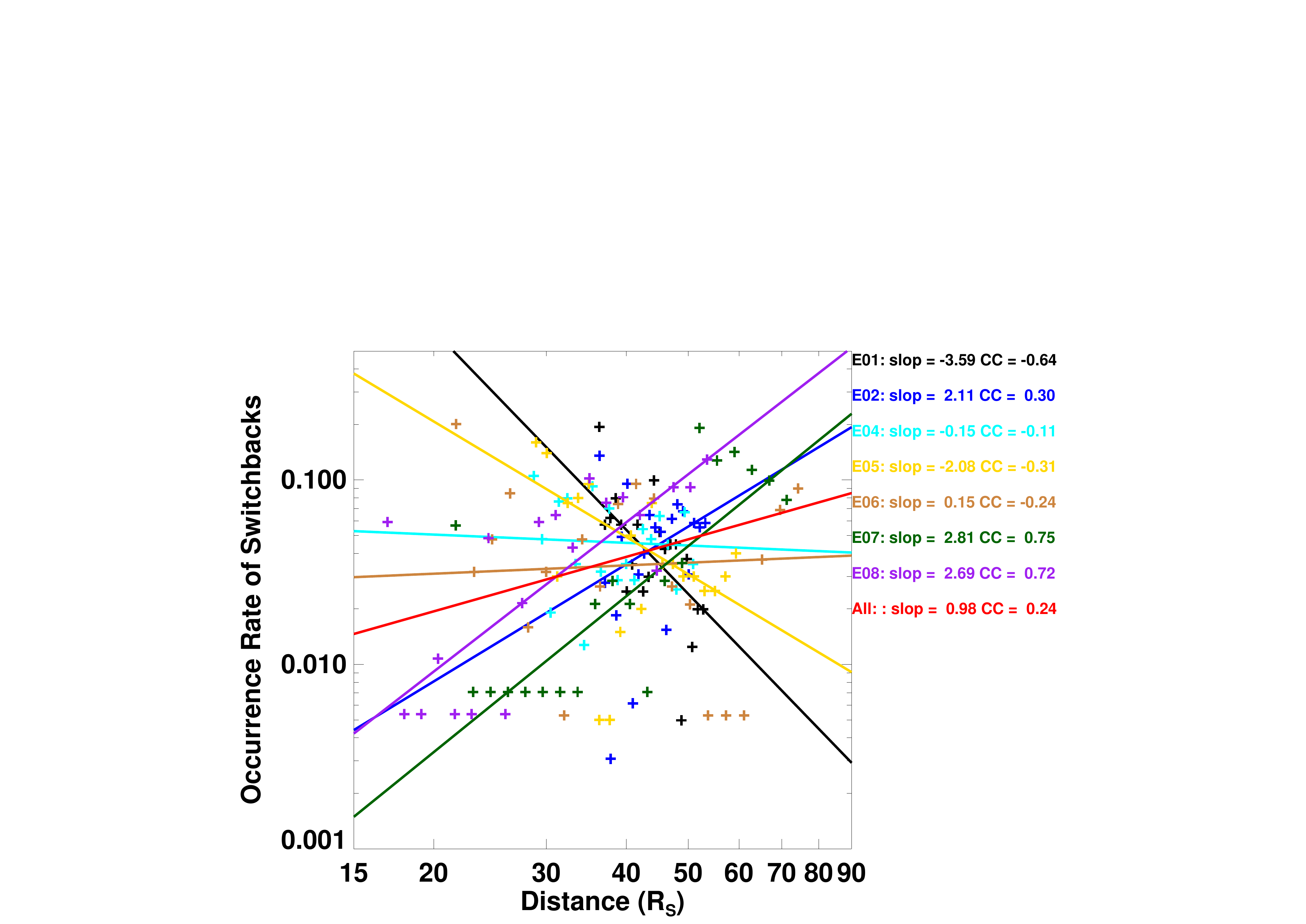}
\caption{The occurrence rates of switchbacks during the seven encounters. The radial variations of the occurrence rate of switchbacks are fitted with a linear function, and the fitted slope and correlation coefficient (CC) during each encounter is presented in the right figure with the same color. }
\label{fig:orate}
\end{figure}

\begin{figure}
\epsscale{0.75}
\plotone{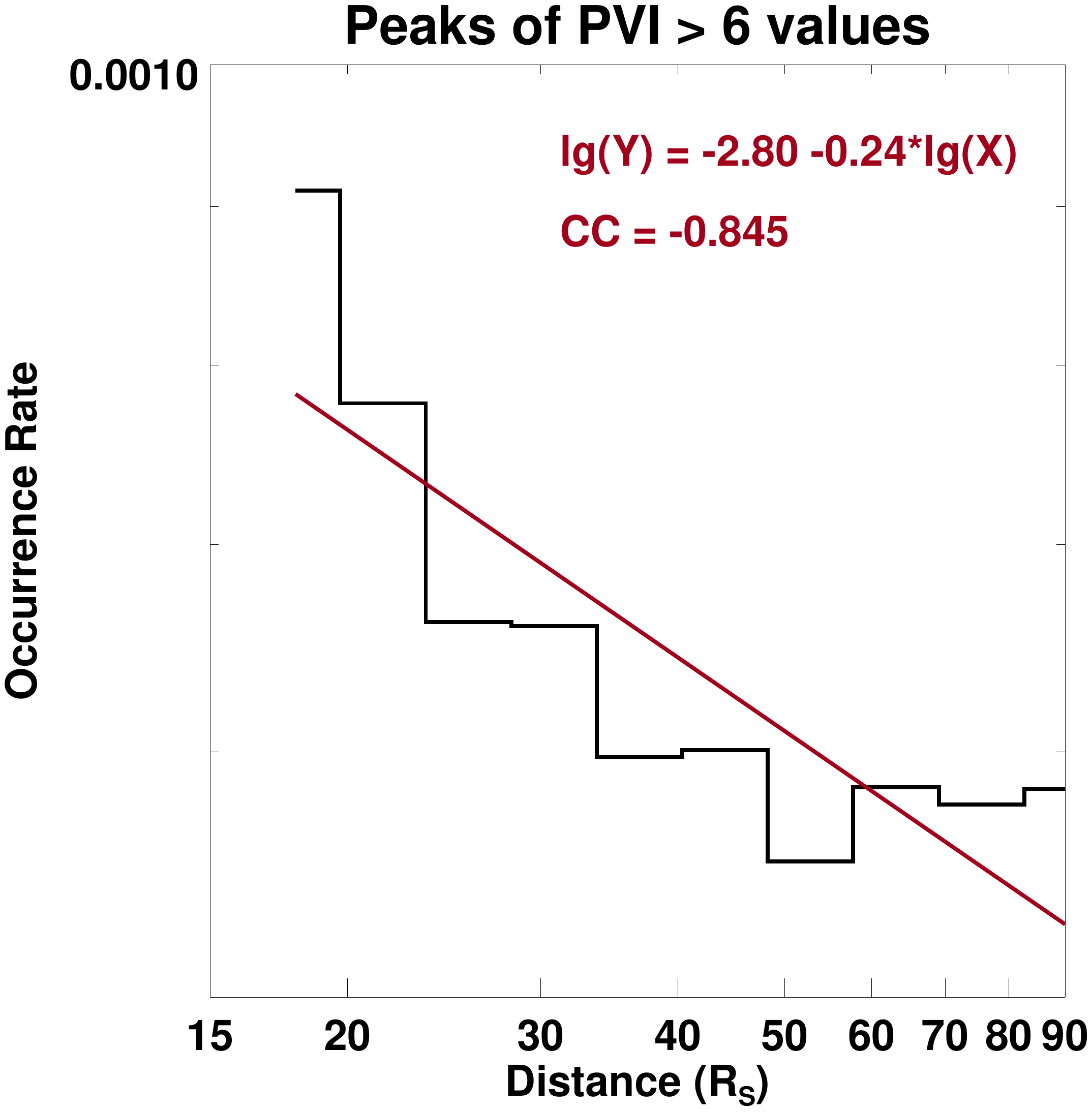}
\caption{The occurrence rate of small-scale current sheets as a function of distance from the Sun. The black histogram indicates the occurrence rate of current sheets at different distances, and the radial evolution is fitted by the red line, with the relationship presented in the upper right of the figure. For the relationship, lg(Y) and lg(X) represent the base 10 logarithm values of occurrence rate and distance, respectively, and CC means the correlation coefficient between the two parameters. }
\label{fig:pvievo}
\end{figure}
The origins of switchbacks are still debatable. According to the introduction, current observations tend to support the solar origins of switchbacks through interchange reconnection processes near the Sun \citep{drake-2020, fisk-2020, woodham-2020, zank-2020}, but other formation methods cannot be excluded. 

The presence of small-scale current sheets in switchbacks is consistent with the idea that switchbacks are formed by interchange reconnection near the Sun. Interchange reconnection occurs between open magnetic flux and large coronal loops. Large coronal loops are stable structures and the randomly braided field lines contribute to the stability of the original loops. Since the braiding is not easy to unwind, it is also not easy to unwind during the interchange reconnection process, and the resulting switchback has inherent stability built in from its beginning. The small-scale current sheets are more abundant inside switchbacks could be a reasonable consequence of the braided magnetic field. As the switchbacks propagate farther, the relaxation of braided magnetic field lines starts from the quiet region and spreads to the transition region and spike. Due to the relatively long durations of transition regions, the spike may dissipate at last, but the relaxation of the braided field lines and the dissipation of current sheets will finally destroy some switchbacks. Consequently, we may further compare the switchback characteristics with this scenario. \citet{Mozer-2020} find the Poynting flux increases significantly inside switchbacks. This is consistent with magnetic braiding, which stores energy in the braided field \citep{ng-2012, yeates-2014}. Moreover, the braiding implies the flux tubes are tilted strongly with respect to each other \citep{borovsky-2008}, which may lead to the large angular deflections of switchbacks \citep{horbury-2020} and the Poynting flux enhancements with rotational angle \citep{Mozer-2020}. Additionally, \citet{larosa-2021} shows an interesting result on the distribution of discontinuities, they find that most switchbacks have a tangential discontinuity at the leading edge and a rotational discontinuity at the trailing edge, or vice versa. This special distribution may be in line with our scenario. In general, the complex topology of the braided field leads to the formation of tangential discontinuities \citep{parker-1972, janse-2010}, while the interchange reconnection could generate rotational discontinuities \citep{lee-1996, lin-2009}. Thus, a switchback that forms from interchange reconnection should result in rotational discontinuity at the reconnection side whilst retaining the tangential discontinuity at the other side. Besides, the "patch" distribution is a distinct signature of switchbacks \citep{horbury-2020, woolley-2020}. We may conceivably connect the "patch" switchbacks to the flux concentrations caused by the footpoint fragmentations at the edges of granules or supergranules \citep{schrijver-1998, berger-2009, berger-2010}, but we should carefully compare their time scales to check the consistency \citep{horbury-2020, bale-2021, fargette-2022}.

In addition, the fact that switchbacks occur in all types of solar winds further implies that they should come from all of their source regions in the corona \citep{fisk-2020}. The helium signatures of switchbacks also support these conclusions. We find both helium-rich and helium-poor populations inside and outside switchbacks, indicating their origins from both closed and open magnetic field regions in the Sun. The $\mathrm{A_{He}}$ - $\mathrm{\Delta V_{\alpha p}/V_A}$ distributions further suggest that the reconnection mechanism should dominate the formation processes, but the wave-turbulence driven mechanism should also contribute. Moreover, both the distributions of small-scale current sheets and the helium signatures show slight asymmetries between the leading and trailing edges of switchbacks, implying the so-called super-Parker spiral \citep{schwadron-2021} may have a limited contribution as the magnetic field footpoints walk from the source of slow wind to faster wind may implement measurable asymmetries between the leading and trailing edges.

However, we could not exclude the possibility that the switchbacks may form in interplanetary space. 
\citet{macneil-2020} give a piece of evidence that the switchbacks observed by Helios increase with distances from 0.3 au to 1 au. They also note that the switchbacks are shorter than 40 s due to the data time resolution, and the occurrence rates at different distances are calculated with data samples, which may bring deviations due to solar wind expansions. Based on our database, 46.6\% of the spikes are shorter than 40 s, thus these small switchbacks could significantly change the radial variations of switchbacks. In order to reduce the influences of solar wind expansions, which could happen as the durations of switchbacks seem to increase with heliocentric distances (not shown), we calculate the occurrence rates at different distances with the number of switchbacks at this distance to the total number of switchbacks in each encounter. As shown in Figure \ref{fig:orate}, the occurrence rates of switchbacks during each encounter are presented with different colored plus signs, and the corresponding radial evolution relationships are shown on the right side. It shows that the occurrence rates decrease with distance during E1 and E4-E5, but increase in E2 and E6-E8, implying that some switchbacks are probably formed in interplanetary space. We note that the fitted relationships could change if we select a different number of distance bins and the occurrence rates are somewhat scattering; we suppose this is related to the "patch" distributions of switchbacks \citep{horbury-2020} and/or the different solar wind conditions in different encounters. If we use the data samples to calculate the occurrence rates, then we can see the rates also show both increase and decrease trends in different encounters, suggesting the switchback durations indeed affect. In addition, the calculation of occurrence rates does not consider the solar wind conditions and the difference in the observation time at different distances. As PSP collects more data in future orbits, we can then extend this analysis with fewer uncertainties. 

Moreover, turbulence may also work to generate switchbacks as suggested by several works \citep[e.g.][]{bourouaine-2020, shoda-2021, wu-2021, pecora-2022}, but it may not be the primary formation mechanism based on our results. PVI values also measure the intermittency in turbulence, thus large PVI values indicate strong intermittency in the turbulence. If the switchbacks are formed by turbulence, then we expect to observe more switchbacks at places where large PVI values are more abundant. Figure \ref{fig:pvievo} displays the occurrence rate of current sheets as a function of distance from the Sun, suggesting the current sheets decrease with the distance and approach to a steady state after about 40 $R_S$. Consequently, if the above assumption is correct, we should see a decreasing trend of switchback occurrence rate with distance. However, Figure \ref{fig:orate} indicates the switchback occurrence rate can even increase beyond 40 $R_S$, which implies that turbulence could not be the primary formation mechanism of switchbacks. We note the latest work of \citet{pecora-2022} found the switchback number per correlation length also increases with distance, and they interpreted the enhancement of switchback presence as the turbulent cascade takes place at a large correlation length. But this also contrasts with the decreasing trend of switchbacks in E1 and E4-E5 as shown in Figure \ref{fig:orate}. However, we agree that turbulence could contribute, to some extent, to the formation of switchbacks in interplanetary space, and further work is needed to determine the prioritized formation mechanism of switchbacks both in the Sun and in the interplanetary medium.

\subsection{Other concerns \label{sec:others}}
Focusing on the inverted magnetic field lines, current studies argue that they could be manifestations of appropriate crossings of magnetic field lines by spacecraft \citep{fisk-2020, macneil-2020}. On one hand, the switchback could be an intact structure that there is no abrupt change of the solar stream during the crossing, i.e. the switchback is formed by the same flux tubes, which is supported by current observations \citep{Mozer-2020, martinovic-2021, woolley-2020, mcmanus-2022} and by our results that the $\mathrm{A_{He}}$ distributions are similar in different regions of switchbacks. On the other hand, the folded magnetic field lines could be formed by the propagation of ejecta or small transients \citep{drake-2020, macneil-2020}, thus the spacecraft may measure the structured solar wind at one side and the intact solar stream from the Sun at the other side. In this circumstance, the solar streams in the switchbacks, especially in LTR and TTR, could be very different, and thus bring bias to trace their origins and add asymmetry between leading and trailing edges. In future works, we may need to separate these switchbacks, and comparisons of the solar wind characteristics (such as the currents, proton entropy, distributions of suprathermal electrons, temperature anisotropies \citep{huang-2020}, etc.) in different regions of switchbacks are necessary to figure out whether a switchback is formed by the same bundle of flux tubes.

\section{Summary} \label{sec:sum}
In this work, we study the stability of switchback structures by investigating the distributions of small-scale current sheets with the PVI method. Our results show that the current sheets are more abundant in spike than both transition regions and quiet periods, and slight asymmetry seems to be present in the leading and trailing edges of switchbacks. Moreover, the occurrence rate of current sheets appears to decrease with radial distances from the Sun, suggesting they may work together with the S-shaped magnetic field curvature of switchbacks to unite the switchback structures and enable them to propagate outward to be observed at least in PSP space. 

The presence of small-scale current sheets in the switchbacks also supports the idea that the switchbacks are formed through interchange reconnection processes, and some observational characteristics of switchbacks are consistent with this scenario. With the alpha measurements available in E4-E8, we analyze the $\mathrm{A_{He}}$ variations inside and outside switchbacks. It shows that both helium-rich and helium-poor populations are observed in different regions of switchbacks, indicating that the switchbacks come from both open and closed magnetic field regions in the Sun. Furthermore, the joint variations of $\mathrm{A_{He}}$ versus $\mathrm{\Delta V_{\alpha p}/V_A}$ further denote that low $\mathrm{A_{He}}$ and low $\mathrm{\Delta V_{\alpha p}/V_A}$ population dominates in different regions of switchbacks, implying magnetic reconnection is the primary mechanism to produce the switchbacks. However, we cannot exclude the possibility that some switchbacks may originate from interplanetary space via other formation mechanisms. 

\begin{acknowledgements}

Parker Solar Probe was designed, built, and is now operated by the Johns Hopkins Applied Physics Laboratory as part of NASA’s Living with a Star (LWS) program (contract NNN06AA01C). Support from the LWS management and technical team has played a critical role in the success of the Parker Solar Probe mission.
Thanks to the Solar Wind Electrons, Alphas, and Protons (SWEAP) team for providing data (PI: Justin Kasper, BWX Technologies). Thanks to the FIELDS team for providing data (PI: Stuart D. Bale, UC Berkeley). J. H. is also supported by NASA grant 80NSSC23K0737. L. K. J. is supported by LWS research program.  

\end{acknowledgements}

\bibliography{switchback}{}
\bibliographystyle{aasjournal}

\end{document}